\documentclass[11pt]{article}
\usepackage{}
\usepackage{amsfonts}
\usepackage{amsmath}
\usepackage{mathrsfs}
\usepackage{amsmath, amssymb}
\usepackage{mathbbold}
\usepackage{latexsym}
\usepackage{CJK}                     
\newtheorem{lemma}{Lemma}[section]

\newtheorem{theorem}{Theorem}[section]

\numberwithin{equation}{section} 

\begin{document}
\begin{CJK*}{GBK}{song}

\title{\bf A Random Weighting Approach for Posterior Distributions
\thanks{This study was supported by the National Natural Science Foundation of China (No. 11071137).}}

\author{\textsc{ Zai-Ying Zhou\thanks{  E-mail address: lizzying@mail.ahnu.edu.cn}}\\[-1pt]
( \textit{\small College of Mathematics and Computer Science,}
\\ \textit{Anhui Normal University, Wuhu 241000,
PR China })\\[-2pt]   }
\date{}
\maketitle


\begin{center}
\begin{minipage}[c]{14cm}
\mbox{} \noindent\textbf{Abstract} In Bayesian theory, calculating a
posterior probability distribution is highly important but usually
difficult. Therefore, some methods have been put forward to deal with such
problem, among which, the most popular one is the asymptotic
expansions for posterior distributions. In this paper, we propose an
alternative method, named random weighting method,  for scaled
posterior distributions, and give an ideal convergence
speed, which serves as the theoretical guarantee for methods of numerical simulations.\\
\mbox{} \noindent\textbf{Keywords} posterior distribution;
asymptotic expansion; random weighting method.\\
\mbox{} \noindent\textbf{Mathematics Subject Classification (2010)} Primary 62E20; Secondary 62F15.\\
\end{minipage}
\end{center}\vskip .1in

\section{Introduction}

Bayesian inference has made a
great progress in the past forty years, including theoretical
researches and practical applications in many fields. The dispute between Bayesian and frequentist schools, about how to choose prior distributions, is
almost disappearing. Now the main issues of Bayesian statistics are
determining the prior probability distributions and
calculating the posterior probability distributions, of which the
latter one is more important. For the comprehensive reviews on the
Bayesian statistics, see the monographs of Lindley (1972), Dey
and Rao (2006), and the references therein.

Let $X=(X_1,\cdots,X_n)$ be an i.i.d. sample from a univariate
distribution $F(x,\theta)$, $\theta\in \Theta$, where $\Theta$ is an
open interval on $\mathcal {R}$ and $\theta$ is regarded as an unknown
parameter to be estimated. In Bayesian statistical inference, we
assume that $\theta$ is random with a prior distribution, say
$v(\theta)$. After taking a simple random sample $X$ from the
conditional distribution $F(x|\theta)$ of $X$ given $\theta$,
Bayesian analysis is carried out based on the posterior distribution
$F(\theta|X)$ given $X$. For the parametric family $F(x,\theta)$
indexed by parameter $\theta$ and for a given prior, the posterior
distribution $F(\theta|X)$ can be obtained explicitly merely when
the prior has a specific form or has a simple form, such as a
conjugate prior or an uninformative one. Generally, the calculation
of the posterior distribution is difficult and involved. To handle
this problem, some researchers have studied asymptotic expansions for
posterior distributions, see Johnson (1967), (1970), Ghosh et al.
(1982), or approximate calculations, see Mao and Tang (1989),
Tierney et al. (1986), and some suggested numerical solutions.

In this paper, we provide an alternative method to deal with  the
problem of approximating the posterior distributions. In terms of
the idea of the so-called random weighting method (Zheng, 1987), we
propose a random weighting method to approximate the standardized
posterior distributions. We prove that the approximation possesses the
ideal convergence speed. The merit of random weighting method can
approximate the sampling distribution in frequentist statistical
analysis. The construction of the random weighting approximation for
the posterior will be postponed to the Section 2. In terms of the
approximate theorem in section \ref{sec:2}, we can understand the
approximation of the posterior distribution from frequency point of
view. This makes it possible to calculate the posterior distribution
without iteration like Monte Carlo Markov Chain technique. The
essence of the random weighting method is to construct random
weights which may sample from the various Dirichlet distributions.
To get better approximation of the distribution, we need to design
different weighting method for different purposes. For detailed investigations on the
choice of the random weight under some specific situation, we
refer to the monograph of Shao and Tu (1996) and the references therein.

The paper is organized as follows. We present the regularity
conditions and the main results in section \ref{sec:2}. And as the
conclusion, we present some discussion in section \ref{sec:4}. The proofs
of the main theorems are given in the Appendix. 

\section{Conditions and the Main Results\label{sec:2}}


First of all, to make the problems under study meaningful, the
following regularity conditions are needed.

Let $X_1,\cdots,X_n$ be i.i.d. random variables having a common c.d.f.
$F(x,\theta)$, $\theta\in \Theta$, where $\Theta$ is an open
interval on $\mathcal {R}$. Suppose $f(x,\theta)$ is the p.d.f. of
$F(x,\theta)$ with respect to some $\sigma$-finite measure $\mu$.
Assume that the parameter $\theta$ has a prior, whose p.d.f is
$\nu(\theta)$ with respect to the Lebesgue measure. A closed interval
$[a_0,b_0]\subset \Theta$, with $a_0<b_0$, satisfies that

\[
\nu(\theta)=
\begin{cases} >0,\quad \ \ &
\theta\in\big(a_0,b_0\big),\\
=0, \quad \ \ & \theta\in\big(a_0,b_0\big)^c.
\end{cases}
\]

Let $c<a_0,d>b_0$, such that $[c,d]\subset \Theta$. Further, we need
the following assumptions:

 A1. For each $\theta\in [c,d]$, $f(x,\theta)$ is measurable with respect to $x$.

 A2. For each $\theta, \theta' \in [c,d]$, where $\theta \neq
 \theta'$, we have $\int|f(x,\theta)-f(x,\theta')|\mu(dx)>0$.

 A3. For each $x$, $f(x,\theta)$ is four times continuously
 differentiable in $\theta \in [c,d]$.

 A4. For each $\theta \in (c,d)$, there exits a neighborhood $U_\theta$ of
 $\theta$, and a measurable function $M_{i\theta}(x)$, ($i=1, 2, 3, 4$),
 such that
 $$\sup_{\theta' \in U_\theta}E_{\theta'}\left|\frac{d^i}{d \theta^i}\log f(x,\theta' )\right|^{4+\alpha}<\infty, \quad i=0,1, 2, 3, 4,$$
$$\left|\frac{d^i}{d \theta^i}\log f(x,\theta')-\frac{d^i}{d \theta^i}\log f(x,\theta'')\right|\leq |\theta'- \theta'' |M_{i\theta}(x),$$
and
$$\sup_{\theta' \in U_\theta}E_{\theta'}(M_{i\theta'}(x))^{3+\alpha}<\infty, $$
where $\alpha$ is a positive constant and $i=1,2,3,4.$

Besides, for any $\theta \in (c,d)$, we assume that
$$I(\theta)=E_{\theta}(-\frac{d^2}{d \theta^2}\log f(x,\theta))>0, $$
$$E_{\theta}(\frac{d}{d \theta}\log f(x,\theta))=0, $$
with $I(\theta)$ being continuous on $[c,d]$. And for any $\theta \in [c,d]$, $\theta' \in (c,d)$, there exist neighborhoods $V_\theta$ and $W_{\theta'}$, such that each neighborhood $V\subset V_\theta$ satisfies that
$$\sup_{\sigma  \in W_\theta'}E_{\sigma }|\sup_{\beta \in V}\log f(x,\beta )|^{3+\alpha}<\infty.$$

A5. $\nu(\theta)$ is twice continuously differentiable on $[a_0,b_0]$.

All these assumptions are reasonable and standard (Pfanzagl, 1973;
Ghosh et al, 1982). Now we present our main theorems as follows.

\begin{theorem}\label{theorem1}
Under conditions A1 to A5, we have
\begin{equation}
\lim_{n\rightarrow \infty} \sqrt{n}\sup_{-\infty<y<\infty}|F_n(y)- \Phi(y)-A(y,X_{-n}/\sqrt{n})|=0\qquad a.s. P_\nu,\label{eq:2.1}
\end{equation}
where $X_{-n}=(X_1,\cdots,X_n)$, $F_n(y)=P\{\sqrt{n}(\theta
-\hat{\theta}_n)b\leq y|X_{-n}\}, $ $\Phi(y)=\int_{-\infty}^y
\frac{1}{\sqrt{2\pi}}\exp(-\frac{x^2}{2})dx,$
$\phi(y)=\frac{1}{\sqrt{2\pi}}\exp(-\frac{y^2}{2})$,
$\hat{\theta}_n$ denotes the maximum likelihood estimator (m.l.e.)
of $\theta$, $(-b^2)$ is the value of the second order derivative of
log likelihood function at $\hat{\theta}_n$, $P_\nu $ is the
marginal probability measure of $X_1$ under the prior p.d.f. $\nu $,
and
\begin{eqnarray}
A(y,X_{-n})&=&-\phi(y)[\nu(\hat{\theta}_n)]^{-1}[b^{-3}a_n(\hat{\theta}_n)\nu(\hat{\theta}_n)(y^2+2)+b^{-1}\rho'(\hat{\theta}_n)],\nonumber\\
\nu'(\theta)&=&\frac{d}{d \theta}\nu(\theta ),\nonumber\\
a_n(\theta)&=&\frac{1}{6n}\sum_{j=1}^n\frac{d^3}{d \theta^3}\log f(x_j,\theta).
\label{eq:2.2}
\end{eqnarray}

\end{theorem}

The proof of the theorem above can be derived directly from Theorem
2.1 of Ghosh et al (1982).

In fact, the $F_n(y)$ in Theorem \ref{theorem1} is the standardized
posterior c.d.f. For now, our goal is to construct a random weighing
statistic, whose c.d.f. $F^\star_n(y)$ can approximate $F_n(y)$.

Let
\begin{equation}
k_n(y)=P^\star\{H_n/\bar{H}_n\leq y\},\label{eq:2.3}
\end{equation}
with
\begin{eqnarray}
&&H_n=\sum_{j=1}^n(\alpha_j-\bar{\alpha})V_j,\nonumber\\
&&\bar{H}_n^2=\frac{1}{n(4n+1)}\sum_{j=1}^n(\alpha_j-\bar{\alpha})^2,\label{eq:2.4}\\
&&\alpha_j=\frac{d}{d \theta}\log f(x_j,\hat{\theta}_n),\quad j=1,\cdots,n,
\ \  \bar{\alpha}=\frac{1}{n}\sum_{j=1}^n \alpha_j,\nonumber
\end{eqnarray}
where $(V_1,\cdots,V_n)$ is a random vector with the Dirichlet
distribution $D(4,4,\cdots,4)$, i.e., $V_1+\cdots+V_n=1$, and the
joint p.d.f. of $(V_1,\cdots,V_{n-1})$ is
\begin{equation}
f(x_1,\cdots,x_{n-1})=\frac{\Gamma(4n)}{\Gamma(4)^n}x_1^3 \cdots x_{n-1}^3(1-x_1-\cdots -x_{n-1})^3,\qquad (x_1,\cdots,x_{n-1})\in S_{n-1}, \label{eq:2.5}
\end{equation}
with
$S_{n-1}=\{(x_1,\cdots,x_{n-1}):x_j\geq 0,\sum_{j=1}^{n-1}x_j\leq 1\}$.
$(V_1,\cdots,V_n)$ and $X_{-n}$ are independent, and letters with
asterisks, such as $P^\star$ and $F^\star$ above, denote conditional
probability and conditional c.d.f. given $X_1,\cdots,X_n$. From the
theories of Dirichlet distribution given in Wilks (1962), if
$Z_1,\cdots,Z_n\sim i.i.d. f(x)$, where $f(x)=\frac{2}{\Gamma (4)}(2x)^3e^{-2x}I_{(x>0)}$, then each $Z_i/(\sum_{l=1}^n
Z_l)$,$i=1,\cdots,n$, has a Dirichlet distribution
$D(4,4,\cdots,4)$, thus the statistic $H_n$ can be rewritten as
\begin{equation}
H_n \stackrel{\text{\tiny{D}}}{=} \sum_{j=1}^n(\alpha_i-\bar{\alpha})Z_j/(\sum_{l=1}^n Z_l), \label{eq:2.6}
\end{equation}
Where $\stackrel{\text{\tiny{D}}}{=}$ stands for equality in
distribution. And the $\bar{H}_n^2$ above turns out to be
conditional variance given $X_1, \cdots, X_n$. Hence,
\begin{eqnarray}
P^\star\{H_n/\bar{H}_n\leq y\}
&=&P^\star\{\sum_{i=1}^n(\alpha_i-\bar{\alpha})Z_i/(\sum_{l=1}^n Z_l)\leq y\bar{H}_n\}\nonumber\\
&=&P^\star\{\sum_{i=1}^n B_{in}(y)(Z_i-2)\leq \rho_n(y)\},\label{eq:2.7}
\end{eqnarray}
with
\begin{eqnarray}
&&B_{in}(y)=A_{in}(y)/(\sum_{j=1}^n A_{jn}^2(y))^{1/2},\quad i=1,\cdots,n,\nonumber\\
&& A_{in}(y)=\alpha_i-\bar{\alpha}-y\bar{H}_n,\nonumber\\
&& \rho_n(y)=-2\sum_{i=1}^n B_{in}(y). \label{eq:2.8}
\end{eqnarray}

Denote $\rho_{in}=(Z_i-2)B_{in}$, $i=1,\cdots,n$, then they are a sequence of independent random variables(r.v.s) given $X_{-n}$.

\begin{theorem}\label{theorem2}
Suppose that conditions A1 to A5 are all satisfied, and $H_n$ is
given by (\ref{eq:2.4}), then
\begin{equation}
\lim_{n\rightarrow \infty} \sqrt{n}\sup_{-\infty<y<\infty}|P^\star\{H_n/\bar{H}_n\leq y\}- \Phi(y)+\frac{1}{6}\phi(y)(y^2-1)\frac{\sum_{i=1}^n(\alpha_i-\bar{\alpha})^3}
{(\sum_{i=1}^n(\alpha_i-\bar{\alpha})^2)^{3/2}}|=0\label{eq:2.9}
\end{equation}

holds for almost every sample sequence $X_1,\cdots,X_n$.

Again, let
\begin{equation}
F_n^\star(y)=P^\star\{\omega_n(H_n/\bar{H}_n)\leq y\},\label{eq:2.10}
 \end{equation}
where
\begin{equation}
\omega_n(y)=(y-\beta_n/\sqrt{n})-\beta_n'/\sqrt{n}(y-\beta_n/\sqrt{n})^2+\beta_n'^2/{3n}(y-\beta_n/\sqrt{n})^3,\label{eq:2.11}
 \end{equation}
with
\begin{eqnarray}
&&\beta_n=-2b^{-3}a_n(\hat{\theta}_n)-b^{-1}\nu'(\hat{\theta}_n)/\nu(\hat{\theta}_n)-\frac{1}{6}\sqrt{n}\frac{\sum_{i=1}^n(\alpha_i-\bar{\alpha})^3}
{(\sum_{i=1}^n(\alpha_i-\bar{\alpha})^2)^{3/2}},\nonumber\\
&&\beta_n'=-b^{-3}a_n(\hat{\theta}_n)+\frac{1}{6}\sqrt{n}\frac{\sum_{i=1}^n(\alpha_i-\bar{\alpha})^3}
{(\sum_{i=1}^n(\alpha_i-\bar{\alpha})^2)^{3/2}}.
\label{eq:2.12}
\end{eqnarray}

\end{theorem}

Thus, we can use $F^\star_n(y)$ to approximate $F_n(y)$. We have the
following conclusion on the precision of this approximation.

\begin{theorem}\label{theorem3}
Assume that conditions A1 to A5 are all met, then
\begin{equation}
\lim_{n\rightarrow \infty} \sqrt{n}\sup_{-\infty<y<\infty}|F_n(y)- F_n^\star(y)|=0\label{eq:2.13}
\end{equation}
is true for almost every sample sequence.

\end{theorem}

\section{Concluding remarks\label{sec:4}}

Finally, we point out that the approximate value of a posterior distribution can be calculated by Monte Carlo method, and the
purpose of this paper is to offer a theoretical guarantee for such
simulation. The random approximation method for posterior distribution can be used to deal with more complicated problems involved in the computation of posterior distribution. The advantage of the method of this paper is easily implemented via computer. However, it is difficult to obtain the closed form except some special parametric family. Some extensions of the method of this paper to multivariate situations may be possible and are beyond the score of this paper. Furthermore, we can provide better approximation to posterior moments.

\newpage
\noindent{\bf Appendix: Proofs of Theorems}


In this section, we present the proofs of Theorem \ref{theorem2} and
Theorem \ref{theorem3} in section \ref{sec:2}.

From the definitions of $F^\star_n$ and $k_n$, we can easily obtain
\begin{equation}
F_n^\star(y)=k_n(\omega_n^{-1}(y)),\label{eq:3.1}
 \end{equation}
 where $\omega_n^{-1}(y)\triangleq u_n(y)$ is the inverse function of $\omega_n(y)$,
 which is given by (\ref{eq:2.11}) and strictly monotone on $(-\infty,\infty)$ with
 range $(-\infty,\infty)$.  Following the line of Yao (1988), we can prove that this $u_n$ takes the form
 \begin{equation}
u_n(y)=\frac{1}{\sqrt{n}}\beta_n+y+\frac{1}{\sqrt{n}}\beta'_ny^2(1+o(1)),\label{eq:3.2}
 \end{equation}
where $o(1)$ stands for an infinitesimal as
$\frac{1}{\sqrt{n}}y\rightarrow 0$.

To prove Theorem \ref{theorem2} and Theorem \ref{theorem3}, we need to
asymptotically expand $k_n(y)$.

Notice (\ref{eq:2.7}), together with Theorem 1 of Bai et al. (1985),
we have
\begin{eqnarray}
&&|P^\star\{H_n/\bar{H}_n\leq y\}- \Phi(\rho_n(y))+\frac{1}{6}\phi(\rho_n(y))(\rho_n^2(y)-1)\sum_{j=1}^n B_{jn}^3|\nonumber\\
&&\leq c\{(1+|\rho_n(y)|)^{-3}\sum_{j=1}^n E^*|W_{jn}(\rho_n(y))|^3 
+(1+|\rho_n(y)|)^{-4}\sum_{j=1}^n E^*|Z_{jn}(\rho_n(y))|^4\nonumber\\
&& \quad {}+(1+|\rho_n(y)|)^{-4}n^6(\sup_{|t|\geq \delta _n}\frac{1}{n}\sum_{j=1}^n|V_{jn}(t)|+\frac{1}{2n})^n \}\nonumber\\
&&=c(J_1+J_2+J_3),\label{eq:3.3}
\end{eqnarray}
where $Y_{jn}=\rho_{jn}I_{\{|\rho_{jn}|\leq 1\}},$
$Z_{jn}(\rho_n(y))=\rho_{jn}I_{\{|\rho_{jn}|\leq 1+|\rho_n(y)|\}},$
$W_{jn}(\rho_n(y))=\rho_{jn}I_{\{|\rho_{jn}|> 1+|\rho_n(y)|\}},$
$\delta_n=\frac{1}{12}\left(\sum_{j=1}^n
E^\star|Y_{nj}|^3\right)^{-1},$ $V_{jn}(t)= E^\star
e^{it\rho_{jn}}=\exp\{-2itB_{jn}\}(1-itB_{jn}/2)^{-4},$ with
$j=1,\cdots,n$ and $E^\star$ being the conditional expectation given
$X_1,\cdots,X_n$.

Next, we will prove that
\begin{equation}
\lim_{n\rightarrow \infty} \sqrt{n}\sup_{-\infty<y<\infty}J_i=0,\qquad i=1,2,3,\label{eq:3.4}
\end{equation}
hold for almost every sample sequence $X_1,\cdots,X_n$.

Simple calculation implies that
\begin{equation}
\rho_n(y)=2\sqrt{n}y/\sqrt{4n+1+y^2}.\label{eq:3.5}
\end{equation}

Now, we study the asymptotic prosperity of each term on the right
side of (\ref{eq:3.3}). Begin with the definition of $B_{jn}$, we
get
$$B_{jn}(y)=(\alpha_j-\bar{\alpha}-y\bar{H}_n)/[(1+y^2/(4n+1))^{1/2}(\sum_{j=1}^n(\alpha_j-\bar{\alpha})^2)^{1/2}].$$
Hence,
\begin{eqnarray}
&&\max_{1\leq j\leq n}\sup_{-\infty<y<\infty}|B_{jn}(y)|\nonumber\\
&&{\leq \max_{1\leq j\leq n}|\alpha_j|/(\sum_{j=1}^n(\alpha_j-\bar{\alpha})^2)^{1/2}+|\bar{\alpha}|/(\sum_{j=1}^n(\alpha_j-\bar{\alpha})^2)^{1/2} +\sup_{-\infty<y<\infty}[y^2/n(4n+1)/(1+\frac{1}{4n+1}y^2)]^{1/2}}\nonumber\\
&&{\leq \max_{1\leq j\leq n}|\alpha_j|/\sqrt{n}/[\frac{1}{n}\sum_{j=1}^n(\alpha_j-\bar{\alpha})^2]^{1/2}+
\frac{1}{\sqrt{n}}|\bar{\alpha}|/[\frac{1}{n}\sum_{j=1}^n(\alpha_j-\bar{\alpha})^2]^{1/2}+1/\sqrt{n}.}
\label{eq:3.6}
\end{eqnarray}
Since
\begin{eqnarray}
\frac{1}{\sqrt{n}}\max_{1\leq j\leq n}|\alpha_j|
&=&\frac{1}{\sqrt{n}}\max_{1\leq j\leq n} |\frac{d}{d\theta}\log f(x_j,\hat{\theta}_n)|\nonumber\\
&\leq& \frac{1}{\sqrt{n}}\max_{1\leq j\leq n} |\frac{d}{d\theta}\log f(x_j,\theta)| \nonumber\\
&&{}+\frac{1}{\sqrt{n}}\max_{1\leq j\leq n} |\frac{d}{d\theta}\log f(x_j,\hat{\theta}_n)-\frac{d}{d\theta}\log f(x_j,\theta)| \nonumber\\
&\leq& \frac{1}{\sqrt{n}}\max_{1\leq j\leq n} |\frac{d}{d\theta}\log f(x_j,\theta)|
+\frac{1}{\sqrt{n}}|\hat{\theta}_n-\theta|\max_{1\leq j\leq n} M_{1\theta}(x_j,\theta),
\label{eq:3.7}
\end{eqnarray}
$\frac{d}{d\theta}\log f(x_j,\theta)$, $j=1,\cdots,n$ is a sequence
of  i.i.d. r.v.s, and Condition A4 shows that
$E|\frac{d}{d\theta}\log f(X,\theta)|^4<\infty$, it follows from the
the standard limit theorem that
\begin{equation}
\lim_{n\rightarrow \infty}\max_{1\leq j\leq n}
|\frac{d}{d\theta}\log f(x_j,\theta)|/\sqrt{n}=0\qquad a.s.
\label{eq:3.8}
\end{equation}
Condition A4 also implies that $EM_{1\theta}^3(X,\theta)<\infty$,
and on the conditions A1 to A4, making use of the method suggested
by Pfanzagl (1973), it can be derived that the m.l.e. sequence of
$\theta$, i.e. $\hat{\theta}_n$, is strongly consistent, thus
\begin{equation}
\lim_{n\rightarrow \infty} \frac{1}{\sqrt{n}}|\hat{\theta}_n-\theta|\max_{1\leq j\leq n} M_{1\theta}(x_j,\theta)=0\qquad a.s. \label{eq:3.9}
\end{equation}
(\ref{eq:3.7}) to (\ref{eq:3.9}) give
\begin{equation}
\lim_{n\rightarrow \infty} \frac{1}{\sqrt{n}}\max_{1\leq j\leq n} |\alpha_j|=0\qquad a.s. \label{eq:3.10}
\end{equation}
Besides, we need to prove that
\begin{equation}
\lim_{n\rightarrow \infty} \bar{\alpha}=0\qquad a.s. \label{eq:3.11}
\end{equation}
and
\begin{equation}
\lim_{n\rightarrow \infty} \frac{1}{n}\sum_{j=1}^n(\alpha_j-\bar{\alpha})^2
=E(\frac{d}{d\theta}\log f(X,\theta))^2>0\qquad a.s. \label{eq:3.12}
\end{equation}
The proof of (\ref{eq:3.11}) is presented as follows.

Since
\begin{eqnarray*}
 |\bar{\alpha}|&=&|\frac{1}{n}\sum_{j=1}^n\frac{d}{d\theta}\log f(x_j,\hat{\theta}_n)| \\
&\leq & |\frac{1}{n}\sum_{j=1}^n\frac{d}{d\theta}\log f(x_j,\theta)|+\frac{1}{n}\sum_{j=1}^n|\frac{d}{d\theta}\log f(x_j,\hat{\theta}_n)-\frac{d}{d\theta}\log f(x_j,\theta)| \\
&\leq & |\frac{1}{n}\sum_{j=1}^n\frac{d}{d\theta}\log
f(x_j,\theta)|+ |\hat{\theta}_n-\theta|\frac{1}{n}\sum_{j=1}^n |
M_{1\theta}(x_j,\theta)|,
\end{eqnarray*}
from Condition A4 and strong law of large numbers comes that
$$\lim_{n\rightarrow \infty} \frac{1}{n}\sum_{j=1}^n\frac{d}{d\theta}\log f(x_i,\theta)=E\frac{d}{d\theta}\log f(X,\theta)=0\qquad a.s.$$
as well as
$$\lim_{n\rightarrow \infty} \frac{1}{n}\sum_{j=1}^nM_{1\theta}(x_i,\theta)=EM_{1\theta}(X,\theta)<\infty\qquad a.s.$$
(\ref{eq:3.11}) holds from the fact that $\hat{\theta}_n$ is a sequence of strongly consistent estimators of $\theta$.

Next, we prove (\ref{eq:3.12}), the details are given as below.
\begin{eqnarray}
&&|\frac{1}{n}\sum_{j=1}^n(\alpha_j-\bar{\alpha})^2-E(\frac{d}{d\theta}\log f(X,\theta))^2| \nonumber\\
&\leq& |\frac{1}{n}\sum_{j=1}^n{\alpha_j}^2-E(\frac{d}{d\theta}\log f(X,\theta))^2|+|\bar{\alpha}|^2 \nonumber\\
&\leq& |\frac{1}{n}\sum_{j=1}^n (\frac{d}{d\theta}\log f( x_j,\theta))^2
-E(\frac{d}{d\theta}\log f(X,\theta))^2| \nonumber\\
&+&|\frac{1}{n}\sum_{j=1}^n [\frac{d}{d\theta}\log f( x_j,\theta)-
\frac{d}{d\theta}\log f(x_j,\hat{\theta}_n)][\frac{d}{d\theta}\log
f( x_j,\theta)+\frac{d}{d\theta}\log
f(x_j,\hat{\theta}_n)]|+|\bar{\alpha}|^2,\label{eq:3.13}
\end{eqnarray}
together with (\ref{eq:3.11}) and strong law of large numbers, on the righthand side of (\ref{eq:3.13}), both of the first and the last term converge to $0$ a.s. and the middle term  can be easily proved to be also convergent to $0$ a.s.. Therefore, (\ref{eq:3.11}) and (\ref{eq:3.12}) hold.

(\ref{eq:3.6}), (\ref{eq:3.10}) to (\ref{eq:3.13}) combined together gives
\begin{equation}
\lim_{n\rightarrow \infty} \max_{1\leq j\leq n}\sup_y |B_{jn}(y)|=0\qquad a.s. \label{eq:3.14}
\end{equation}

Now we are ready to prove (\ref{eq:3.4}).

When $i=1$, by strong law of large numbers, similar proof with that of
(\ref{eq:3.11}) and (\ref{eq:3.12}) derives that
\begin{equation}
\lim_{n\rightarrow \infty} \sup_{-\infty<y<\infty} \sqrt{n}\sum_{j=1}^n|B_{jn}|^3<\infty,\qquad a.s.\label{eq:3.16}
\end{equation}
Together with (\ref{eq:3.14}) and (\ref{eq:3.16}), we have
\begin{eqnarray}
&&\lim_{n\rightarrow \infty} \sqrt{n} \sup_{-\infty<y<\infty} J_1
\nonumber\\
&=&\lim_{n\rightarrow \infty} \sqrt{n} \sup_{-\infty<y<\infty}(1+|\rho_n(y)|)^{-3}\sum_{j=1}^n E^\star|W_{jn}(\rho_n(y))|^3 \nonumber\\
&=&\lim_{n\rightarrow \infty} \sqrt{n} \sup_{-\infty<y<\infty}(1+|\rho_n(y)|)^{-3}\sum_{j=1}^n E^\star|\rho_{jn}|^3 I_{\{|\rho_{jn}|> 1+|\rho_n(y)|\}}\nonumber\\
&=&\lim_{n\rightarrow \infty} \sqrt{n} \sup_{-\infty<y<\infty}(1+|\rho_n(y)|)^{-3}\sum_{j=1}^n |B_{jn}|^3 E|Z_j-2|^3 I_{\{|Z_j-2||B_{jn}|> 1+|\rho_n(y)|\}}\nonumber\\
&\leq &\lim_{n\rightarrow \infty} \sqrt{n} \sup_{-\infty<y<\infty}(1+|\rho_n(y)|)^{-3}\sum_{j=1}^n |B_{jn}|^3 E|Z_j-2|^3 I_{\{|Z_j-2|\geq (1+|\rho_n(y)|)/\max_{1\leq j\leq n}|B_{jn}|\}}\nonumber\\
&\leq &\lim_{n\rightarrow \infty} [\sqrt{n} \sup_{-\infty<y<\infty}\sum_{j=1}^n |B_{jn}|^3] E|Z_1-2|^3 I_{\{|Z_1-2|\geq (1+|\rho_n(y)|)/\max_{1\leq j\leq n}|B_{jn}|\}}\nonumber\\
&=& 0 \qquad a.s. \label{eq:3.15}
\end{eqnarray}

When $i=2$, by (\ref{eq:3.14}) and (\ref{eq:3.16}), we get
\begin{eqnarray}
\lim_{n\rightarrow \infty} \sqrt{n} \sup_{-\infty<y<\infty} J_2
&=&\lim_{n\rightarrow \infty} \sqrt{n} \sup_{-\infty<y<\infty}(1+|\rho_n(y)|)^{-4}\sum_{j=1}^n E^*|Z_{jn}(\rho_n(y))|^4 \nonumber\\
&=&\lim_{n\rightarrow \infty} \sqrt{n} \sup_{-\infty<y<\infty}(1+|\rho_n(y)|)^{-4}\sum_{j=1}^n E^*|\rho_{jn} I_{\{|\rho_{jn}|\leq  1+|\rho_n(y)|\}}|^4 \nonumber\\
&\leq &\lim_{n\rightarrow \infty} \sqrt{n} \sup_{-\infty<y<\infty}(1+|\rho_n(y)|)^{-4}\sum_{j=1}^n E^*|\rho_{jn}|^4 \nonumber\\
&\leq &\lim_{n\rightarrow \infty} \sqrt{n} \max_{1\leq j\leq n}\sup_{-\infty<y<\infty} |B_{jn}(y)|\cdot \sum_{j=1}^n|B_{jn}(y)|^3\cdot  E|Z_1-2|^4 \nonumber\\
&=& 0 \qquad a.s. \label{eq:3.17}
\end{eqnarray}

When $i=3$, since
\begin{eqnarray*}
\sum_{j=1}^n E^\star|Y_{jn}|^3&=&\sum_{j=1}^n E^\star|\rho_{jn}|^3 I_{\{|\rho_{jn}|\leq 1\}}\\
&=&\sum_{j=1}^n|B_{jn}|^3 E|Z_1-2| I_{\{|Z_1-2||B_{jn}|\leq  1\}}\\
&\leq & \max_{1\leq j\leq n}\sup_{-\infty<y<\infty} |B_{jn}(y)|\cdot \sum_{j=1}^n B_{jn}^2\qquad (\sum_{j=1}^n B_{jn}^2=1)\\
&=&\max_{1\leq j\leq n}\sup_{-\infty<y<\infty}
|B_{jn}(y)|\rightarrow 0 \qquad a.s.
\end{eqnarray*}
we get
\begin{equation}
\delta_{n}=\frac{1}{12}(\sum_{j=1}^n E^\star|Y_{jn}|^3)^{-1}\rightarrow \infty \qquad a.s.\label{eq:3.18}
\end{equation}
whose convergence is uniform in $y\in (-\infty,\infty)$. Besides,
\begin{eqnarray}
\lim_{n\rightarrow \infty} \sqrt{n} \sup_{-\infty<y<\infty} J_3
&&=\lim_{n\rightarrow \infty} \sqrt{n} \sup_{-\infty<y<\infty} (1+|\rho_n(y)|)^{-4}n^6[\sup_{|t|\geq \delta_n}\sum_{j=1}^n |V_{jn}(t)|+\frac{1}{2n}]^n  \nonumber\\
& &\leq\lim_{n\rightarrow \infty} n^7 \sup_{-\infty<y<\infty}
[\sup_{|t|\geq \delta_n}\frac{1}{n}\sum_{j=1}^n |exp\{-2itB_{jn}\}(1-itB_{jn}/2)^{-4}|+\frac{1}{2n}]^n \nonumber\\
&&=\lim_{n\rightarrow \infty} n^7 \sup_{-\infty<y<\infty} [\sup_{|t|\geq \delta_n}\frac{1}{n}\sum_{j=1}^n (1/(1+t^2 B_{jn}^2/4))^2+\frac{1}{2n}]^n \nonumber\\
&&\leq \lim_{n\rightarrow \infty} n^7 \sup_{-\infty<y<\infty} [\frac{1}{n}\sum_{j=1}^n 1/(1+\delta_n^{\frac{1}{2}} B_{jn}^2)+\frac{1}{2n}]^n \nonumber\\
&&\leq \lim_{n\rightarrow \infty} n^7 \sup_{-\infty<y<\infty} (1-\frac{1}{n}\delta_n^{\frac{1}{2}}+\frac{1}{n}\sum_{j=1}^n |B_{jn}|^4\delta_n+\frac{1}{n})^n, \label{eq:3.19}
\end{eqnarray}
and notice the term in the middle of the right side of (\ref{eq:3.19})
\begin{eqnarray*}
\sum_{j=1}^n |B_{jn}|^4\delta_n &=& \frac{1}{12}\cdot
\frac{\sum_{j=1}^n B_{jn}^4}{\sum_{j=1}^n |B_{jn}|^3 E|Z_j-2|^3I_{\{|B_{jn}(Z_j-2)|\leq 1\}}}\\
&\leq & \max_{1\leq j\leq n}\frac{ |B_{jn}|}{12 E|Z_1-2|^3I_{\{|B_{jn}(Z_1-2)|\leq 1\}}}
\rightarrow 0 \qquad a.s.
\end{eqnarray*}
Meanwhile, it can be derived from axioms of probability
theories that
$$\lim_{n\rightarrow \infty} \sup \sup_{-\infty<y<\infty} \frac{1}{\sqrt{n}}\delta_n<\infty,$$
and
$$\lim_{n\rightarrow \infty} \inf \inf_{-\infty<y<\infty} \frac{1}{\sqrt{n}}\delta_n>0.$$
From these we know that there exits a constant $c>0$, such that when n grows sufficiently large, almost every sample sequence has
$$ (1-\frac{1}{\sqrt{n}}\cdot \frac{1}{\sqrt{n}}\delta_n^{\frac{1}{2}}+\frac{1}{n}+\frac{1}{n}\sum_{j=1}^n |B_{jn}|^4\delta_n)^n<e^{-cn^{1/2}},$$
therefore,
\begin{equation}
\lim_{n\rightarrow \infty}\sqrt{n} \sup_{-\infty<y<\infty} J_3=0 \qquad a.s.\label{eq:3.20}
\end{equation}
 Recall (\ref{eq:3.15}), (\ref{eq:3.17}) and (\ref{eq:3.20}), it can be inferred that (\ref{eq:3.5}) true.

 To summarize what has been mentioned above, we have Lemma  \ref{lemma1} as below:
 \begin{lemma}\label{lemma1}
If conditions A1 to A5 are all satisfied, then for almost every sample sequence $X_1,\cdots,X_n,\cdots$, we have
\begin{equation}
\lim_{n\rightarrow \infty}\sqrt{n} \sup_{-\infty<y<\infty}
|P^\star\{H_n/\bar{H}_n\leq y\}- \Phi(\rho_n(y))+\frac{1}{6}\phi(\rho_n(y))(\rho_n^2(y)-1)\sum_{j=1}^n B_{jn}^3|=0.\label{eq:3.21}
\end{equation}
\end{lemma}

\begin{lemma}\label{lemma2}  (See Tu and Zheng (1987) )    There exits a constant $c>0$, such that
\begin{eqnarray}
\sup_{-\infty<y<\infty} |\Phi(\rho_n(y))-\Phi(y)|\leq c/n, \label{eq:3.22}\\
\sup_{-\infty<y<\infty} |\phi(\rho_n(y))(\rho_n^2(y)-1)-\phi(y)(y^2-1)|\leq c/n. \label{eq:3.23}
\end{eqnarray}
\end{lemma}

The proof of Theorem \ref{theorem2} is as below. Let
$$r_n=\sum_{j=1}^n(\alpha_j-\bar{\alpha})^3/(\sum_{j=1}^n(\alpha_j-\bar{\alpha})^2)^{3/2},$$
$$\bar{r}_n=\sum_{j=1}^n(\alpha_j-\bar{\alpha}-y\bar{H}_n)^3/(\sum_{j=1}^n(\alpha_j-\bar{\alpha}-y\bar{H}_n)^2)^{3/2}.$$
Since
\begin{eqnarray}
&&\sqrt{n} \sup_{-\infty<y<\infty}
|P^\star\{H_n/\bar{H}_n\leq y\}- \Phi(y)+\frac{1}{6}\phi(y)(y^2-1)r_n| \nonumber\\
&\leq& \sqrt{n} \sup_{-\infty<y<\infty} |P^\star\{H_n/\bar{H}_n\leq y\}- \Phi(\rho_n(y))+\frac{1}{6}\phi(\rho_n(y))(\rho_n^2(y)-1)\sum_{j=1}^n B_{jn}^3| \nonumber\\
&&+\sqrt{n} \sup_{-\infty<y<\infty}|\Phi(\rho_n(y))- \Phi(y)|+ \sqrt{n} \sup_{-\infty<y<\infty} |\frac{1}{6}[\phi(y)(y^2-1)-\phi(\rho_n(y))(\rho_n^2(y)-1)]\sum_{j=1}^n B_{jn}^3| \nonumber\\
&&{}+\sqrt{n}\sup_{y>(4n+1)^{1/4}}[|\frac{1}{6}\phi(y)(y^2-1)|(|r_n|+|\sum_{j=1}^n B_{jn}^3|)]\nonumber
\end{eqnarray}
\begin{eqnarray}
&&{}+\sqrt{n}\sup_{y<-(4n+1)^{1/4}}|\frac{1}{6}\phi(y)(y^2-1)|[|r_n|+|\sum_{j=1}^n B_{jn}^3|]\nonumber\\
&&{}+\sqrt{n} \sup_{|y|\leq (4n+1)^{1/4}}|\frac{1}{6}\phi(y)(y^2-1)||r_n-\bar{r}_n|.\label{eq:3.24}
\end{eqnarray}

It can be derived from Lemma \ref{lemma1} and Lemma \ref{lemma2} that the former 3 terms on the right side of (\ref{eq:3.24}) are convergent to $0$ for almost every sample sequence $X_1,\cdots,X_n,\cdots$.

When $y>(4n+1)^{1/4}$ , $ \frac{1}{6}\phi(y)(y^2-1)$ is monotonically decreasing and
$$\lim_{n\rightarrow \infty}\frac{1}{6}\phi((4n+1)^{1/4})((4n+1)^{1/2}-1)=0,$$
with (\ref{eq:3.16}), we know that the fourth term of the right side of (\ref{eq:3.24}) is convergent to $0$ for almost every sample sequence $X_1,\cdots,X_n,\cdots,$. The similar method can be applied on the fifth term.  As for the last term, notice that
$$\sup_{-\infty<y<\infty}|\phi(y)(y^2-1)|<\infty$$
and the definition of $\bar{H}_n$, it is not difficult to prove that
\begin{eqnarray*}
&&\sqrt{n} \sup_{|y|\leq (4n+1)^{1/4}} |r_n-\bar{r}_n|\\
&=&\sup_{|y|\leq (4n+1)^{1/4}}|\frac{\frac{1}{n}\sum_{j=1}^n(\alpha_j-\bar{\alpha})^3}{(\frac{1}{n}\sum_{j=1}^n(\alpha_j-\bar{\alpha})^2)^{3/2}}
-\frac{\frac{1}{n}\sum_{j=1}^n[\alpha_j-\bar{\alpha}-\frac{y}{\sqrt{4n+1} }\cdot \sqrt{\frac{1}{n}\sum_{j=1}^n(\alpha_j-\bar{\alpha})^2}]^3}
{[\frac{1}{n}\sum_{j=1}^n(\alpha_j-\bar{\alpha})^2+\frac{y^2}{4n+1}\cdot \frac{1}{n}\sum_{j=1}^n(\alpha_j-\bar{\alpha})^2]^{3/2}}|\rightarrow 0.
\end{eqnarray*}
Thus the proof of Theorem \ref{theorem2} is complete.

The proof of Theorem \ref{theorem3} is given as below.

(\ref{eq:3.1}) implies that
\begin{equation}
F^*_n(y)=k_n(u_n(y))=P^\star\{H_n/\bar{H}_n\leq u_n(y)\}.\label{eq:3.25}
\end{equation}
And from Theorem \ref{theorem2}, for almost every sample sequence $X_1,\cdots,X_n,\cdots$, we have
\begin{equation}
\lim_{n\rightarrow \infty}\sqrt{n} \sup_{-\infty<y<\infty}
|F^*_n(y)-\Phi(u_n(y))+\frac{1}{6}\phi(u_n(y))(u_n^2(y)-1)r_n|=0.\label{eq:3.26}
\end{equation}
Using (\ref{eq:2.1}) of Theorem \ref{theorem1} gives that
\begin{eqnarray}
&&\lim_{n\rightarrow \infty}\sqrt{n} \sup_{-\infty<y<\infty} |F^*_n(y)-F_n(y)|\nonumber\\
&=&\lim_{n\rightarrow \infty}\sqrt{n} \sup_{-\infty<y<\infty}
|\Phi(u_n(y))-\frac{1}{6}\phi(u_n(y))(u_n^2(y)-1)r_n-\Phi(y)
\nonumber\\
&&+\frac{1}{\sqrt{n}}\phi(y)
[2b^{-3}a_n(\hat{\theta }_n)+b^{-1}\nu'(\hat{\theta }_n)/\nu(\hat{\theta }_n)+b^{-3}a_n(\hat{\theta }_n)y^2]|\nonumber
\end{eqnarray}
\begin{eqnarray}
&\leq & \lim_{n\rightarrow \infty}\sqrt{n} \sup_{|y|\leq n^{\frac{1}{12}}} |\Phi(u_n(y))-\frac{1}{6}\phi(u_n(y))(u_n^2(y)-1)r_n-\Phi(y)-\frac{1}{\sqrt{n}}A(y,X_{-n})|
\nonumber\\
&&+\lim_{n\rightarrow \infty}\sqrt{n} \sup_{y>n^{\frac{1}{12}}} |\Phi(u_n(y))-\frac{1}{6}\phi(u_n(y))(u_n^2(y)-1)r_n-\Phi(y)-\frac{1}{\sqrt{n}}A(y,X_{-n})|
\nonumber\\
&&+\lim_{n\rightarrow \infty}\sqrt{n} \sup_{y<-n^{\frac{1}{12}}} |\Phi(u_n(y))-\frac{1}{6}\phi(u_n(y))(u_n^2(y)-1)r_n-\Phi(y)-\frac{1}{\sqrt{n}}A(y,X_{-n})|
\nonumber\\
&=&I_1+I_2+I_3. \label{eq:3.27}
\end{eqnarray}

First, we deal with $I_1$. The mean value theorem states that there exit points $\xi $ and $\eta $ between $u_n(y)$ and $y$, such that
\begin{eqnarray}
\Phi(u_n(y))-\Phi(y)&=&\phi(y)(u_n(y)-y)+R_{1n}\label{eq:3.28}\\
\phi(u_n(y))(u_n^2(y)-1)-\phi(y)(y^2-1)&=&\phi(y)(3y-y^3)(u_n(y)-y)+R_{2n}
. \label{eq:3.29}
\end{eqnarray}
Recalling (\ref{eq:3.2}), (\ref{eq:3.28}) and (\ref{eq:3.29}) gives that
\begin{eqnarray}
I_1&=&\lim_{n\rightarrow \infty}\sqrt{n} \sup_{|y|\leq n^{\frac{1}{12}}} |\Phi(u_n(y))-\Phi(y)-\frac{1}{6}\phi(u_n(y))(u_n^2(y)-1)r_n-\frac{1}{\sqrt{n}}A(y,X_{-n})|
\nonumber\\
&=&\lim_{n\rightarrow \infty}\sqrt{n} \sup_{|y|\leq n^{\frac{1}{12}}} |\phi(y)(u_n(y)-y)+R_{1n}-\frac{1}{6}[\phi(u_n(y))(u_n^2(y)-1)-\phi(y)(y^2-1)]r_n\nonumber\\
&&-\frac{1}{6}\phi(y)(y^2-1)r_n- \frac{1}{\sqrt{n}}A(y,X_{-n})|
\nonumber \\
&=&\lim_{n\rightarrow \infty}\sqrt{n} \sup_{|y|\leq
n^{\frac{1}{12}}} |
\frac{1}{\sqrt{n}}\phi(y)y^2[-b^{-3}a_n(\hat{\theta }_n)+\frac{1}{6} \sqrt{n}r_n]o(1)+R_{1n}\nonumber\\
&&-\frac{1}{6}\phi(y)(3y-y^3)(u_n(y)-y)r_n+R_{2n}r_n|
\nonumber\\
&\leq &\lim_{n\rightarrow \infty} \sup_{|y|\leq
n^{\frac{1}{12}}} |\phi(y)y^2||b^{-3}a_n(\hat{\theta
}_n)+\frac{1}{6}\sqrt{n}r_n|
\sup_{|y|\leq n^{\frac{1}{12}}} |o(1)|\nonumber\\
&&+\lim_{n\rightarrow \infty}\sqrt{n}\sup_{|y|\leq n^{\frac{1}{12}}} |R_{1n}|
+\lim_{n\rightarrow \infty}\sup_{|y|\leq n^{\frac{1}{12}}}
\frac{1}{6}|\phi(y)(3y-y^3)||(u_n(y)-y)||\sqrt{n}r_n|
+\lim_{n\rightarrow \infty}\sup_{|y|\leq n^{\frac{1}{12}}} |R_{2n}|\sqrt{n}r_n\nonumber\\
&=&I_{11}+I_{12}+I_{13}+I_{14}. \label{eq:3.30}
\end{eqnarray}
Use Conditions A4, A5 and strong law of large numbers, it is easy to check that
\begin{equation}\label{eq:3.31}
  \left.\begin{aligned}
         \lim_{n\rightarrow \infty}|b^{-3}a_n(\hat{\theta }_n)|&<\infty \qquad a.s.\\
        \lim_{n\rightarrow \infty}\sqrt{n}r_n&<\infty \qquad a.s.\\
       \lim_{n\rightarrow \infty}|b^{-1}\nu'(\hat{\theta }_n)/\nu(\hat{\theta }_n)|&<\infty \qquad a.s.
        \end{aligned}
  \qquad \right\}
\end{equation}
which can be either proved in the same way of that of (\ref{eq:3.11}) and (\ref{eq:3.12}) or directly derived from Lemma 4 of \cite{3}.

When $|y|\leq n^{\frac{1}{12}}$, $y/\sqrt{n}\rightarrow 0$, and it follows from the definition of $o(1)$ that
\begin{equation}\label{eq:3.32}
\lim_{n\rightarrow \infty}\sup_{|y|\leq n^{\frac{1}{12}}} |o(1)|=0.
\end{equation}
Notice that there exits a constant $k>0$, such that
\begin{equation}\label{eq:3.33}
\sup_{-\infty <y<\infty} |y^i\phi(y)|\leq k<\infty,\quad i=0,1,\cdots,4.
\end{equation}
Therefore, it holds from (\ref{eq:3.31}) to (\ref{eq:3.33}) that
\begin{equation}\label{eq:3.34}
I_{11}=0.
\end{equation}

(\ref{eq:3.2}) shows that
\begin{eqnarray}
I_{12}&=&\lim_{n\rightarrow \infty}\sqrt{n}\sup_{|y|\leq n^{\frac{1}{12}}} |R_{1n}|\nonumber\\
&=&\lim_{n\rightarrow \infty}\sqrt{n} \sup_{|y|\leq n^{\frac{1}{12}}}
|\frac{1}{2}\xi\phi(\xi)||u_n(y)-y|^2 \nonumber\\
&\leq &k \lim_{n\rightarrow \infty}\sqrt{n} \sup_{|y|\leq n^{\frac{1}{12}}}|\frac{1}{\sqrt{n}}\beta_n+\frac{1}{\sqrt{n}}\beta'_n y^2(1+o(1))|^2 \nonumber \\
&\leq &2k \lim_{n\rightarrow \infty} (\frac{1}{\sqrt{n}}|\beta_n|^2+|\beta'_n|^2 \cdot \sup_{|y|\leq n^{\frac{1}{12}}} \frac{1}{\sqrt{n}}y^4(1+|o(1)|) )\nonumber\\
&\leq &0+2k \lim_{n\rightarrow \infty}
|\beta'_n|^2\cdot\frac{1}{\sqrt{n}} n^{\frac{4}{12}} (1+\sup_{|y|\leq n^{\frac{1}{12}}} |o(1)|)
=0. \label{eq:3.35}
\end{eqnarray}

Repeating the same argument gives
\begin{equation}
I_{13}=0, \ \ I_{14}=0.\label{eq:3.36}
\end{equation}
(\ref{eq:3.34}) to (\ref{eq:3.36}) imply that
\begin{equation}
I_{1}=0.\label{eq:3.37}
\end{equation}

Concerning $I_{3}$, notice that $\phi(y)|y|^i,(i=0,1,2)$ are all monotonically increasing on $(-\infty,2)$ with
$$\lim_{y\rightarrow \infty}\phi(y)|y|^i=0,\qquad i=0,1,2.$$
And use (\ref{eq:3.31}), we have
\begin{eqnarray*}
I_3&\leq& \lim_{n\rightarrow \infty} \sqrt{n}\Phi(u_n(-n^{1/12}))+\frac{1}{6}\lim_{n\rightarrow \infty}
 \phi(u_n(-n^{1/12}))(|u_n(-n^{1/12})|+1)|\sqrt{n}r_n|\\
&&+ \lim_{n\rightarrow \infty} \sqrt{n}\Phi(-n^{1/12})+\lim_{n\rightarrow \infty}\phi(-n^{1/12})[|2b^{-3}a_n(\hat{\theta }_n)|+|b^{-1}\nu'(\hat{\theta }_n)/\nu(\hat{\theta }_n)|] \\
&&+ \lim_{n\rightarrow \infty}|b^{-3}a_n(\hat{\theta }_n)|\phi(-n^{1/12})\cdot n^{\frac{1}{12}}  \\
&=& \lim_{n\rightarrow \infty} \sqrt{n}\Phi(u_n(-n^{1/12}))\\
&=&  \lim_{n\rightarrow \infty} \sqrt{n}\Phi(\frac{1}{\sqrt{n}}\beta _n-n^{1/12}+\frac{1}{\sqrt{n}}\beta'_n n^{1/6}(1+o(1)))\\
&\leq &\lim_{n\rightarrow \infty} \sqrt{n}\Phi(-n^{1/12}+1)=0,
\end{eqnarray*}
hence
\begin{equation}
I_{3}=0.\label{eq:3.38}
\end{equation}

As for $I_{2}$, the fact of $\phi(y)|y|^i,(i=0,1,2)$ all being monotonically increasing on $(2,\infty)$ gives that
\begin{eqnarray*}
I_2&\leq& \lim_{n\rightarrow \infty} \sqrt{n} (1-\Phi(u_n(n^{1/12})))
+\lim_{n\rightarrow \infty} \sqrt{n} (1-\Phi(n^{1/12}))\\
&&+\frac{1}{6}\lim_{n\rightarrow \infty}
 \phi(u_n(n^{1/12}))|u_n(n^{1/12})||\sqrt{n}r_n|\\
&&+\lim_{n\rightarrow \infty}
\phi(-n^{1/12})[|12b^{-3}a_n(\hat{\theta }_n)|+|b^{-1}\nu'(\hat{\theta }_n)/\nu(\hat{\theta }_n)|] \\
&&+\lim_{n\rightarrow \infty}
|b^{-3}a_n(\hat{\theta }_n)\phi(n^{1/12})n^{1/12} \\
&=&\lim_{n\rightarrow \infty} \sqrt{n}(1-\Phi(u_n(n^{1/12})))
+\lim_{n\rightarrow \infty} \sqrt{n} (1-\Phi(n^{1/12}))\\
&=&  0,
\end{eqnarray*}
that is
\begin{equation}
I_{2}=0.\label{eq:3.39}
\end{equation}

Then Theorem \ref{theorem3} follows right from (\ref{eq:3.37}) to (\ref{eq:3.39}).

\def\refname

\end{CJK*}

\enddocument